\documentclass[10pt,conference]{IEEEtran}
\IEEEoverridecommandlockouts

\usepackage[inline,shortlabels]{enumitem}
\usepackage{mathtools}
\usepackage{braket}
\usepackage{quantikz}
\usepackage{caption}
\usepackage{subcaption}
\usepackage{adjustbox}
\usepackage{comment}

\usepackage{cite}
\usepackage{amsmath,amssymb,amsfonts}

\usepackage[bookmarks=true,breaklinks=true,letterpaper=true,colorlinks,citecolor=blue,linkcolor=blue,urlcolor=blue]{hyperref}

\newcommand{\hh}[1]{}

\usepackage[ruled,noend]{algorithm2e}
\SetKwInOut{Input}{Input}
\SetKwInOut{Output}{Output}
\SetKw{Break}{break}
\SetKw{Or}{or}
\usepackage{etoolbox}
\AtBeginEnvironment{algorithm}{\small}

\DeclareMathOperator\dist{dist}

\SetKwComment{Comment}{\#\,}{}
\definecolor{mygreen}{HTML}{004a17}

\SetCommentSty{mycommfont}

\def\BibTeX{{\rm B\kern-.05em{\sc i\kern-.025em b}\kern-.08em
    T\kern-.1667em\lower.7ex\hbox{E}\kern-.125emX}}
\begin{document}

\newcommand{\brand}{TeleSABRE}
\newcommand{\ie}{\textit{i.e.}}
\newcommand{\symArch}{\mathcal{A}}
\newcommand{\symCirc}{\mathcal{G}}

\title{\brand{}: Layout Synthesis in Multi-Core Quantum Systems with Teleport Interconnect}

\iftrue
\author
{\IEEEauthorblockN{Enrico Russo, Elio Vinciguerra, Maurizio Palesi, Davide Patti, Giuseppe Ascia, Vincenzo Catania}
\IEEEauthorblockA{
Department of Electrical, Electronic, and Computer Engineering (DIEEI), University of Catania, I-95125 Catania, Italy.\\ \textit{enrico.russo@phd.unict.it, elio.vinciguerra@phd.unict.it, \{name.surname\}@unict.it}
}
}
\else
\IEEEspecialpapernotice{(Omitted for blind review)}
\fi

\maketitle
\begin{abstract}
  Quantum circuit compilation and, in particular, efficient qubit layout synthesis is a critical challenge in modular, multi-core quantum architectures with constrained interconnects. In this work, we extend the SABRE heuristic algorithm to develop \brand{}, a layout synthesis approach tailored for architectures featuring teleportation-based interconnects. Unlike standard SABRE, which only introduces SWAP operations for qubit movement, \brand{} integrates both intracore SWAPs and teleportation-based techniques leveraging qubit teleportation and gate teleportation across cores. This enables more efficient circuit execution by reducing both inter-core communication overhead and the number of intra-core SWAPs required to allow teleportation protocols and local gate executions. Experimental results demonstrate that \brand{} achieves $28\%$ reduction across various benchmarks in terms of inter-core operations while also taking into account the logistics of the teleport protocols.
\end{abstract}

\section{Introduction}

Quantum computing offers a different approach to computation, with remarkable potential for addressing intricate problems that remain intractable for classical computing systems. Quantum computing promises exponential speedups in domains such as cryptography, molecular simulation, optimization, and machine learning, leveraging quantum mechanical phenomena such as superposition and entanglement to process information \cite{montanaro2016quantum, shor1999polynomial,grover1996fast,nielsen2010quantum,preskill2018quantum,aaronson2013quantum}.

\begin{figure}
    \centering
    \includegraphics{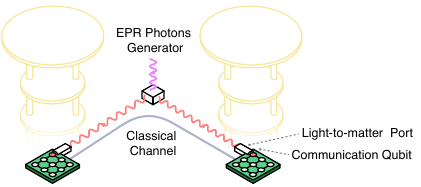}
    \caption{Example multi-core superconducting quantum computer diagram with EPR mediated inter-core communication.}
    \label{fig:computer}
\end{figure}

To realize this potential, quantum computing systems must scale beyond current limitations. However, fundamental physical constraints in quantum processor design impose significant challenges on qubit density and interconnectivity within a single quantum processing unit (QPU). The obstacles to scaling include nanoscale fabrication challenges, stringent cryogenic requirements, extensive control electronics and wiring footprints, crosstalk and undesired interactions between densely packed qubits, and reduced manufacturing yield as chip complexity increases \cite{jnane2022multicore,xie2021mitigating, sarovar2020detecting, arute2019quantum}. These factors collectively make it increasingly difficult to scale single quantum processor chips without compromising qubit fidelity and operational reliability.
As shown in Fig.~\ref{fig:computer}, a promising approach to overcome these scaling limitations involves the development of modular multi-core quantum computing architectures, where multiple smaller QPUs are interconnected to form a larger, more powerful quantum computing system~\cite{smith2022scaling,alarcorn_iscas23}. 

Quantum compilation is responsible for transforming abstract quantum algorithms into executable operations on physical hardware. A significant challenge in this process stems from the limited connectivity between physical qubits in most current quantum architectures. These QPUs typically restrict interactions to physically adjacent qubits. This architectural constraint necessitates the introduction of SWAP operations, \ie{} quantum gates that exchange the states of two physical qubits to enable interactions between logically connected but physically distant qubits. SWAP gates increase circuit depth, \ie{} computation timesteps, potentially causing decoherence, introduce additional noise, and reduce overall fidelity. Consequently, efficient SWAP insertion and routing strategies have become central research problems in quantum compilation, with significant implications for quantum algorithm performance. Optimizing these operations requires sophisticated techniques that balance the trade-offs between circuit depth, gate count, and the specific connectivity topology of target hardware platforms.

The compilation problem for single-core systems is NP-complete~\cite{botea2018complexity, siraichi2018qubit} but the complexity increases dramatically in emerging multi-core quantum systems, where the compilation problem extends beyond intra-core routing to include inter-core communication. These modular architectures can rely on quantum teleportation protocols to transfer quantum information between physically separated processing units or to execute gate on quantum states on two different cores. Unlike conventional SWAP operations, teleportation protocols require the consumption of entangled resource states, classical communication channels, and introduce latency and preparation steps considerations that must be carefully managed. Compilers for such distributed quantum systems must not only optimize the placement and routing within individual cores but also orchestrate the timing and resources for inter-core operations.

The SABRE heuristic~\cite{sabre} has previously addressed qubit routing challenges in single core quantum architectures. This paper introduces \brand{}, a quantum routing framework designed for multi-core quantum processors that extends SABRE heuristics. By utilizing quantum teleportation protocols, \brand{} enables efficient qubit transfer and remote gate applications across quantum processing cores. The framework systematically optimizes three critical metrics: the number of qubit swaps, quantum teleportation operations, and remote telegate applications.
The core technical contribution of \brand{} is its routing algorithm, which employs Dijkstra's shortest path method within the SABRE framework to develop a graph-based approach to quantum resource allocation. This method enables precise qubit routing across multiple quantum processing cores, addressing the specific connectivity constraints of multi-core quantum processor architectures. The reference implementation of the proposed method is publicly available: \url{https://github.com/Haimrich/telesabre}.

This work is organized as follows: Section~\ref{sec:background} provides background on quantum circuits, compilation, and multi-core quantum architectures with teleportation-based interconnects. Section~\ref{sec:problem} formally defines the layout synthesis problem in this context. Section~\ref{sec:methodology} presents the proposed TeleSABRE heuristic, detailing its algorithmic design and energy model. Section~\ref{sec:experiments} evaluates the performance of TeleSABRE against state-of-the-art and near-optimal methods. Section~\ref{sec:limitations} discusses limitations and potential future directions, while Section~\ref{sec:related_works} reviews related work. Finally, Section~\ref{sec:conclusions} concludes the paper.

\section{Background}
\label{sec:background}



\subsection{Quantum Circuits}

\begin{figure}
    \centering
    \begin{subfigure}[b]{\linewidth}
    \scalebox{0.85}{
        \begin{quantikz}[row sep=0.2cm, column sep=0.35cm]
            \lstick{$q_0$} & & & & \ctrl{2} & & & & \ctrl{2} & \ctrl{1} & \gate{T} & \ctrl{1} & \\
            \lstick{$q_1$} & & \ctrl{1} & & & & \ctrl{1} & \gate{T} & & \targ{} & \gate[disable auto height]{T^\dag} & \targ{} & \\
            \lstick{$q_2$} & \gate{H} & \targ{} & \gate[disable auto height]{T^\dag} & \targ{} & \gate{T} & \targ{} & \gate[disable auto height]{T^\dag} & \targ{} & \gate{T} & \gate{H} & &
        \end{quantikz}
    }
    \caption{Circuit diagram.}
    \label{fig:circuit}
    \end{subfigure}
    \begin{subfigure}[b]{\linewidth}
    \includegraphics[width=\linewidth]{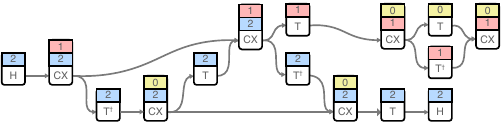}
    \caption{Corresponding direct acyclic graph.}
    \label{fig:circuit_dag}
    \end{subfigure}
    \caption{The quantum circuit representing a decomposed Toffoli gate.}
\end{figure}

In the gate-based model of quantum computation, quantum programs are expressed as sequences of reversible quantum gates, denoted as $G$. These quantum programs, commonly referred to as quantum circuits, manipulate quantum information through operations on logical qubits. An example of a quantum circuit is illustrated in Fig.~\ref{fig:circuit}. The gates within a quantum circuit act on qubits, thereby altering their quantum states.

The state of a qubit is represented as a linear superposition of basis states:
\begin{equation}
    \ket{\phi} = \alpha\ket{0} + \beta\ket{1} = \alpha \begin{bmatrix} 1 \\ 0 \end{bmatrix} + \beta \begin{bmatrix} 0 \\ 1 \end{bmatrix} = \begin{bmatrix} \alpha \\ \beta \end{bmatrix},
\end{equation}
where $\alpha, \beta \in \mathbb{C}$. The states $\ket{0}$ and $\ket{1}$ form a basis for a two-dimensional complex vector space, implying that $\ket{\phi} \in \mathbb{C}^2$. A quantum gate is represented by a unitary matrix that operates on a qubit, thereby modifying its state. For instance, the Hadamard gate $H$ is a commonly used single-qubit gate, defined as:
\begin{equation}
    \ket{\phi}' = H\ket{\phi} = \frac{1}{\sqrt{2}}\begin{bmatrix} 1 & 1 \\ 1 & -1 \end{bmatrix}\ket{\phi}.
\end{equation}

Quantum circuits typically involve multiple logical qubits, with their collective state represented as the tensor product of individual qubit states, assuming the absence of entanglement. For two qubit states $\ket{\phi_1}$ and $\ket{\phi_2}$, the overall system state $\ket{\psi}$ is given by:
\begin{equation}
\begin{split}
    \ket{\psi} &= \ket{\phi_1} \otimes \ket{\phi_2} = 
    \begin{bmatrix} \alpha_1 & \beta_1 \end{bmatrix}^\top
    \otimes
    \begin{bmatrix} \alpha_2 & \beta_2 \end{bmatrix}^\top
    \\ 
    &=  \alpha_1\alpha_2\ket{00} + \alpha_1\beta_2\ket{01} + \alpha_2\beta_1\ket{10} + \alpha_2\beta_2\ket{11}.
\end{split}
\end{equation}
In general, the quantum state of an $n$-qubit system resides in the Hilbert space $\mathbb{C}^{2^n}$. Consequently, a quantum gate operating on $m$ qubits is represented by a $2^m \times 2^m$ unitary matrix, where $m \leq n$.

Although an infinite set of quantum gates can be mathematically represented as unitary matrices, only a finite subset is physically realizable on quantum hardware. The set of gates implemented on a quantum processor constitutes a \textit{universal gate set}, meaning that any quantum circuit can be decomposed into operations using only these gates. Typically, quantum hardware supports single-qubit and two-qubit gates, with the Controlled-NOT (CX) gate being the most common two-qubit gate. The CX gate, denoted as $\text{CX}_{q_1,q_2}$, flips the state of the target qubit $q_2$ if and only if the control qubit $q_1$ is in the $\ket{1}$ state.

\subsection{Circuit Compilation}

\begin{figure}
    \centering
    \includegraphics[width=\linewidth]{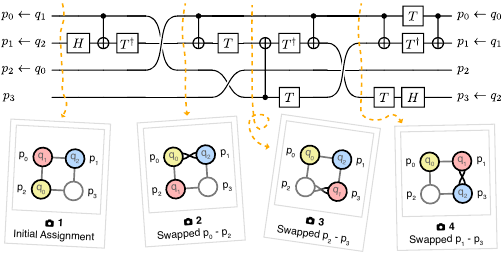}
    \caption{Compiled circuit diagram with additional SWAPs.}
    \label{fig:swaps}
\end{figure}

To execute a quantum circuit, it is necessary first to transform it into an equivalent circuit composed of gates available in the hardware. Additionally, circuit \textit{logical qubits} must be mapped onto \textit{physical qubits}, \ie{} physical devices that behave as two-state quantum systems. The spatial arrangement of physical qubits is particularly significant for two-qubit gate operations, as such gates require direct physical connectivity between the participating qubits. In sparsely connected monolithic quantum hardware architectures, logical qubits must often be repositioned across different physical qubits through SWAP gates to facilitate gate execution. Fig.~\ref{fig:swaps} shows an example compilation for the decomposed Toffoli gate circuit (Fig.~\ref{fig:circuit}). Initially the logical qubits are assigned to the four physical qubit of a four physical qubit architecture. This initial allocation allows the execution of the first $CX_{q_1,q_2}$ gate of the original circuit. However, this initial layout does not allow the execution of the second $CX_{q_0,q_2}$ gate because $q_0$ and $q_2$ are allocated on $p_2$ and $p_1$ respectively, hence they are not adjacent. For this reason, SWAP gates are introduced to make sure that every time a two-qubit gate has to be applied on logical qubits, the physical qubits hosting them are connected. These additional SWAP operations introduce fidelity loss and contribute to execution noise, thereby impacting overall circuit performance.

\subsection{Multi-Core Quantum Computing}

\begin{figure}
    \centering
    \includegraphics[width=\linewidth]{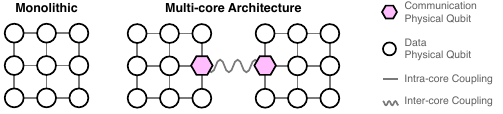}
    \caption{Monolithic and multi-core architecture physical qubits coupling graph comparison.}
    \label{fig:multicore}
\end{figure}

Multi-core quantum computing represents an architectural paradigm in which multiple quantum processing units (QPUs) are interconnected within a single quantum computing system. This architecture, also known as a Quantum Network-on-Chip (QNoC), enables enhanced scalability by mitigating the qubit connectivity constraints inherent in monolithic quantum processors.

A multi-core quantum system consists of multiple tightly integrated QPUs, where the qubits within each core can be efficiently manipulated using local quantum gates. Inter-core quantum operations, however, require specialized mechanisms due to the limitations imposed by quantum mechanics, such as the no-cloning theorem and the challenges associated with quantum state transfer. These interconnects are realized through various physical means, including direct qubit interactions, photonic links, or coherent quantum buses.

One of the primary advantages of multi-core quantum computing is its reduced dependency on long-range quantum communication, making it a more feasible near-term approach compared to distributed quantum computing. The spatial proximity of interconnected cores enables faster qubit interactions while minimizing decoherence and error rates. Additionally, this architecture facilitates the separation of qubits into distinct functional categories, such as computational qubits for processing and memory qubits with extended coherence times for quantum information storage.

Despite its advantages, multi-core quantum computing faces significant challenges. Efficient inter-core communication mechanisms must be developed to ensure high-fidelity quantum state transfer while minimizing noise and gate errors. Moreover, the architecture requires optimized qubit allocation strategies to balance workload distribution and maintain computational efficiency. Future advancements in quantum interconnect technologies and error correction methods will play a crucial role in the scalability and practicality of multi-core quantum architectures.

\subsection{Inter-core Communication}
\label{sec:teleport}

In multi-core quantum architectures, it may be necessary to perform operations between logical qubits located in different cores. To facilitate such remote operations, appropriate communication primitives must be implemented~\cite{caleffi2024distributed}. Unlike classical distributed and modular systems, where information can be freely copied and transferred, quantum information is subject to the no-cloning theorem and must therefore be physically transported between cores. While direct qubit transfer using photons is theoretically feasible, it suffers from low success rates due to attenuation and decoherence. To address these limitations, quantum entanglement is leveraged to enable communication protocols based on quantum teleportation, which allow for the transmission of a qubit without requiring its physical movement. In this work, we consider two fundamental communication primitives for inter-core interconnection, both of which rely on the quantum teleportation protocol. For a more comprehensive treatment of this topic, the interested reader is referred to~\cite{caleffi2024distributed}. Both communication primitives require two entangled quantum states distributed between the communicating cores. This distribution can be accomplished for example using an EPR photon generator as illustrated in Fig.\ref{fig:computer}. The physical qubits that can receive and maintain these entangled quantum states are designated as \textit{communication qubits} while the others allowing only intra-core operations are named \textit{data qubits}, as depicted in Fig.\ref{fig:multicore}.

\subsubsection{Teledata}

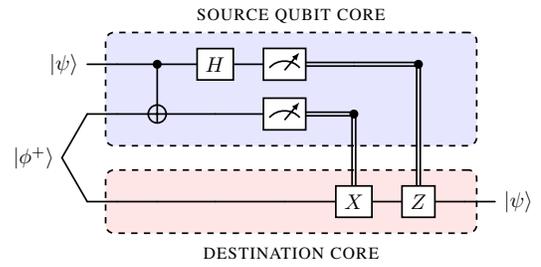
\begin{figure}
    \centering
    \scalebox{0.8}{
    \begin{quantikz}[row sep=0.3cm]
 \lstick{$\ket{\psi}$} & \gategroup[2,steps=7,style={dashed,rounded
corners,fill=blue!10, inner
xsep=2pt},background,label style={label
position=above,anchor=north,yshift=0.3cm}]{{\sc
source qubit core}} & \ctrl{1} & \gate{H} & \meter{} & \setwiretype{c} & \ctrl[vertical wire=c]{2} \\
    \makeebit[angle=-30]{$\ket{\phi^+}$} & & \targ{} && \meter{} & \ctrl[vertical wire=c]{1} \setwiretype{c}  \\[4ex]
     & \gategroup[1,steps=7,style={dashed,rounded
corners,fill=red!10, inner
xsep=2pt},background,label style={label
position=below,anchor=north,yshift=-0.3cm}]{{\sc
destination core}} & & & & \gate{X} & \gate{Z} & & \rstick{$\ket{\psi}$} 
\end{quantikz}
}
    \caption{Qubit teleportation protocol realizing teledata primitive.}
    \label{fig:teledata}
\end{figure}

The \textit{teledata} communication primitive enables the teleportation of a qubit from one core to another by leveraging quantum entanglement. Prior to teleportation, a maximally entangled pair of qubits also known as Bell state is distributed to communication qubits of source and destination cores. As shown in Fig.~\ref{fig:computer}, one way to achieve the distribution of the Bell state is employing a device generating two entangled photons that are transmitted to the two cores and interact with the communication qubits through a light-to-matter port~\cite{rodrigo2021modelling}. \mbox{$\ket{\Phi^+} = \left(\ket{01}+\ket{10}\right) / \sqrt{2}$} is an example of such a state, in this case if a communication qubit is measured as $1$ the other one will collapse to $0$ and viceversa. To transmit the state of a data qubit, it is first entangled with the local communication qubit at the source core, then both qubits are measured. The outcome of this measurement is then classically communicated to the destination core, where the corresponding correction operations are applied to the remote communication qubit, effectively reconstructing the original data qubit. In particular if the output of communication qubit measurement is one, a Pauli-X gate is applied on destination core communication qubit; then, if the output of data qubit measurement is one, a Pauli-Z gate is applied on destination core communication qubit. A schematic circuit representation of this protocol is provided in Fig.~\ref{fig:teledata}. 


\subsubsection{Telegate}

\begin{figure}
    \centering
    \scalebox{0.8}{
    \begin{quantikz}[row sep=0.3cm]
 \lstick{$\ket{\psi_A}$} & \gategroup[2,steps=7,style={dashed,rounded
corners,fill=blue!10, inner
xsep=0pt},background,label style={label
position=above,anchor=north,yshift=0.3cm}]{{\sc
control qubit core}} & \ctrl{1} & & & & \gate{Z} & & \\
    \makeebit[angle=-30]{$\ket{\phi^+}$} & & \targ{} && \meter{} & \ctrl[vertical wire=c]{2} \setwiretype{c} \\[4ex]
     & \gategroup[2,steps=7,style={dashed,rounded
corners,fill=red!10, inner
xsep=0pt},background,label style={label
position=below,anchor=north,yshift=-0.3cm}]{{\sc
target qubit core}} & \ctrl{1} & \gate{H} & \meter{} & \setwiretype{c} & \ctrl[vertical wire=c]{-2} \\
\lstick{$\ket{\psi_B}$} &  & \targ{} & & & \gate{X}& & &
\end{quantikz} = \begin{quantikz}[wire types={q,n,n,q}, column sep=0.2cm]
&\ctrl{3}& \ghost{H}\\
&\ghost{H} &  \\
&\ghost{H} & \\
&\targ{} &  \ghost{H}
\end{quantikz}
}

    \caption{Quantum teleportation protocol realizing telegate primitive.}
    \label{fig:telegate}

\end{figure}
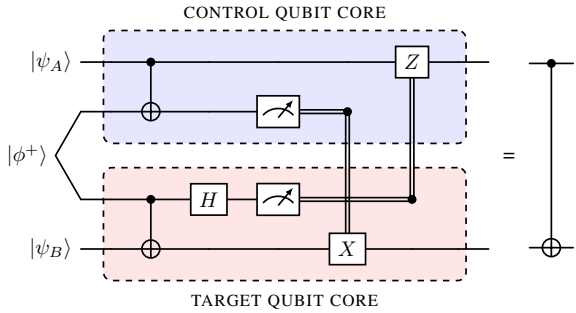

The \textit{telegate} primitive enables the execution of a quantum gate between logical qubits allocated on physical qubits in different cores without requiring their movement. Similar to the teledata primitive, the distribution of a Bell state between the two processors is required. The operation is realized as follows: \begin{enumerate*}[(i)]
    \item two local CX gates are performed in the two cores, one between each data qubit and its corresponding communication qubit
    \item in the target core, an Hadamard gate is applied on the communication qubit
    \item on each core a conditional operation on the data qubit, determined by the measurement outcome of the remote communication qubit
\end{enumerate*}. 
As shown in the corresponding quantum circuit diagram in Fig.~\ref{fig:telegate}, the telegate emulates a direct CX gate between two data qubits in different cores.

\section{Problem Statement}
\label{sec:problem}

\begin{figure*}
    \centering
    \includegraphics[width=\linewidth]{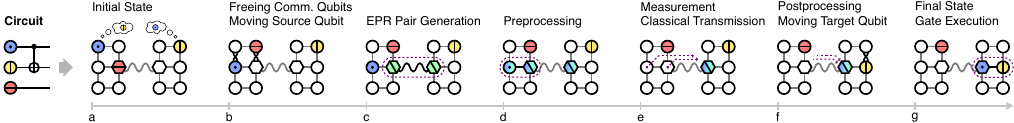}
    \caption{Example multi-core quantum layout synthesis scenario.}
    \label{fig:problem}
\end{figure*}
Consider the scenario depicted in Fig.~\ref{fig:problem}, in which a gate must be executed between two logical qubits that are initially located in different cores (Fig.~\ref{fig:problem}a). The compiler must determine a sequence of additional operations that either: (1) move the two interacting logical qubits to physical positions that enable execution of the telegate operation, or (2) relocate both logical qubits to the same core and then to physical qubits that allow local gate execution. In Fig.~\ref{fig:problem}, we assume the compiler selects the latter approach.

If the compiler's long-term strategy involves performing a teledata operation, the allocation of logical qubits in physical communication qubits must be carefully managed. In Fig.~\ref{fig:problem}b, for instance, it becomes necessary to relocate the logical qubit initially assigned to the communication qubit, thereby freeing it for subsequent teleportation operations. Following this, as required by the teledata protocol (Fig.~\ref{fig:teledata}), a Bell state can be distributed to the two communication qubits (Fig.~\ref{fig:problem}c). The protocol then proceeds with the preprocessing step (Fig.~\ref{fig:problem}d), measurement and classical communication, i.e., the transmission of two classical bits, also known as the ``phone call'' (Fig.~\ref{fig:problem}e), and conditional gate execution (Fig.~\ref{fig:problem}f).

It is important to note that traditional SWAP insertions needed in single-core architectures must also be considered to enable gate execution in the destination core. Consequently, in the step shown in Fig.~\ref{fig:problem}f, a SWAP operation is applied to facilitate local gate execution in the final step (Fig.~\ref{fig:problem}g).

Similar considerations apply to telegate execution. Furthermore, since a circuit typically consists of multiple gates, optimizing the number of qubit movement operations (SWAPs, telegates, and teleports) requires comprehensive analysis to select operations that benefit the greatest number of gates to be executed. When choosing between telegate and teleport operations, the compiler should consider the frequency of future interactions involving logical qubits currently allocated in each core. For more complex multi-core architectures, inter-core routing becomes necessary to determine the optimal teleportation path between cores. All these factors must be balanced while also addressing intra-core movements required for local gate execution and for freeing communication qubits needed in teleportation protocols.

\section{Methodology}
\label{sec:methodology}

In this section, we introduce the \brand{} heuristic algorithm. In Section~\ref{sec:overview} we describe the top-level flow of the proposed algorithm mutuated from SABRE~\cite{sabre}. Then, in Section~\ref{sec:total_energy} we provide details about the energy calculation and we distinguish between local energy calculation (Sec.~\ref{sec:local_energy}) and remote energy calculation (Sec.~\ref{sec:remote_energy}). Finally, we describe the possibility to further optimize the starting layout in Sec.~\ref{sec:initial} and we highlight the algorithmic complexity in Sec.~\ref{sec:complexity}.

\subsection{Algorithm Overview}
\label{sec:overview}

\begin{algorithm}

{\small
\caption{SABRE-based Search Algorithm}\label{alg:overview}
\KwIn{Circuit DAG $\symCirc{}$, Architecture $\symArch{}$, $seed$}
\KwOut{Routed Circuit, Final Layout $\lambda_f$}

$i \gets 0$\;
$\lambda_0 \gets$ \texttt{initialLayout($\symArch{}$, $seed$)}\;
$usage \gets \texttt{onesArray}(\symArch.num\_qubits$)\;
 \While{\texttt{front}($\symCirc{}$) $\neq \varnothing$}{
 $executable\_gates = \varnothing $ \; 
  \For{$gate \in \texttt{front}(\symCirc{})$}{
      \If{\texttt{canRunGate($\symArch{}$,$\lambda_i$,$gate$)}}{
         $executable\_gates.\texttt{append}(gate) $\;
    }
  }
   \eIf{$executable\_gates \neq \varnothing$}{
   \For{$gate \in executable\_gates$}{
        $\symCirc{}.\texttt{removeNode}(gate)$\;
    }
    $\lambda_{i+1} = \lambda_{i}$\;
    }{
    $scores = []$\;
    $candidate\_ops = \texttt{ObtainCandidateOps}(\symArch{}, \symCirc{}, \lambda_i)$\;
    \For{$op \in candidate\_ops$}{
    $\lambda_{op} = \texttt{applyOperation}(\lambda_{i}, op)$\;
    $energy = \texttt{calculateEnergy}(\symArch{}, \symCirc{}, \lambda_{i})$\;
    $scores[op] = energy * \max \{usage[p] \mid p \in op\}$\;
    }
    $op \gets \texttt{sample}(\{op \mid scores[op] = \min scores\})$\;
    
    $\lambda_{i+1} = \texttt{applyOperation}(\lambda_{i}, op)$\;
    $usage \gets \texttt{updateUsagePenalties}(usage, op)$\;
    }
    $i \gets i + 1$\;
}
}
\end{algorithm}

The primary flow of the algorithm, based on SABRE~\cite{sabre}, is outlined in Algorithm~\ref{alg:overview}. \brand{} takes as input a circuit directed acyclic graph (DAG), denoted as $\symCirc{}$ (Fig.~\ref{fig:circuit_dag}), and a multi-core quantum architecture, $\symArch{}$, and produces a sequence of layout-feasible gate executions along with additional SWAP and teleportation operations.

The algorithm begins by generating an initial layout, which assigns logical qubits to physical qubits as described in Sec.~\ref{sec:initial}. Subsequently, the main loop of the algorithm is initiated. In each iteration, the algorithm considers the front layer of the remaining DAG, which consists of the remaining gates eligible for execution, i.e., the subset of gates whose dependencies have been satisfied. The gates in this frontier that are executable given the current qubit layout are scheduled for execution and subsequently removed from the DAG. If no gates are available for execution, the algorithm considers introducing a movement operation to facilitate progress.

In the original SABRE algorithm, the only available movement operation is the SWAP operation. In contrast, \brand{} extends this approach by incorporating additional movement operations, specifically logical qubit teleportation and gate teleportation (telegate). Each candidate operation is evaluated based on a heuristic score function, which we refer to in this work as energy. The algorithm virtually applies each candidate operation to the current state and computes the energy of the resulting layout. The operation that yields the resulting layout with lowest energy is selected for execution. In cases where multiple candidates result in the same energy, one is randomly selected.

To ensure load balancing and minimize circuit depth, the algorithm incorporates a physical qubit usage factor into the energy calculation called decay. This factor is updated each time an operation is applied to a physical qubit, promoting an efficient distribution of movement operation across the architecture inreasing parallelization and reducing compiled circuit depth.

\subsection{Energy Calculation}
\label{sec:total_energy}

In the original SABRE~\cite{sabre,zou2024lightsabre} implementation, the energy of a given state is computed as follows:  
\begin{equation}\label{eq:sabre}
    \text{energy} =
    {\frac{1}{\lvert F\rvert} \sum_{g \in F} \text{energy}^G_\lambda(g)}
    + \underbrace{\frac{k}{\lvert H\rvert} \sum_{(i,j) \in H} \text{energy}^G_\lambda(g)}_{\text{lookahead component}}
\end{equation}
where $F = \texttt{front}(\symCirc{})$ represents the current frontier of the circuit DAG, consisting of gates whose dependencies have been satisfied and are ready for execution. The set $H$, referred to as the \textit{extended set}, serves as a lookahead horizon, encompassing a subset of gates from the remaining circuit DAG that are scheduled for execution in subsequent steps. The parameter $k$ denotes the lookahead factor, which adjusts the relative importance of future gates. The function $\text{energy}^G_\lambda(g)$ represents the energy associated to the gate $g$ given the current qubit layout $\lambda$.

\subsection{Local Energy Calculation}
\label{sec:local_energy}

In the original SABRE algorithm, for single core quantum architectures, the energy associated to each two-qubit gate $g$ acting on logical qubits $q_{g,1}$ and $q_{g,2}$ is calculated as:
\begin{equation}
    \text{energy}^G_\lambda(g)=\dist_\lambda(g)=\mathbf{D}[\text{phys}_\lambda(q_{g,1})][\text{phys}_\lambda(q_{g,2})]
\end{equation}
where $\text{phys}_\lambda(q_{g,1})$ and $\text{phys}_\lambda(q_{g,2})$ are physical qubits hosting $q_{g,1}$ and $q_{g,2}$ respectively according to layout $\lambda$, and $\mathbf{D}$ is the hop distance matrix for physical qubits in the quantum architecture.

The application of a SWAP operation can either increase or decrease this potential energy: it increases if it results in a greater overall distance between pairs of interacting qubits (i.e., those involved in the same gate), and conversely, it decreases if it brings these qubits closer together.  

The distance matrix can be precomputed in $O(N^3)$ time using the Floyd-Warshall algorithm for a given coupling map of a monolithic quantum processor with $N$ physical qubits. This precomputed matrix is then utilized throughout the execution of SABRE to determine the distance between interacting logical qubits.  

\subsection{Remote Energy Calculation}
\label{sec:remote_energy}

\begin{figure*}
    \centering
    \includegraphics[width=\linewidth]{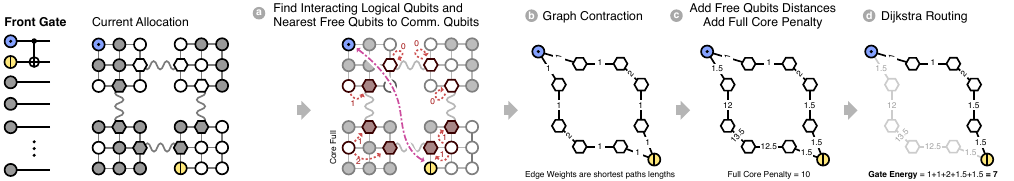}
    \caption{Contracted graph generation and remote gate energy calculation.}
    \label{fig:contracted}
\end{figure*}

Extending this approach to a teleportation-interconnected multi-core architecture, however, is non-trivial. As described in Section~\ref{sec:teleport}, executing a teleportation protocol is a multi-step process. Simply considering the distance between interacting qubits, even if they reside in different cores, is insufficient. A teleportation operation requires several intermediate steps: first, two communication qubits must be entangled; then, the source qubit must be entangled with the mediator physical qubit. Only after these steps can the quantum state be transferred to the target communication qubit. From a layout perspective, a teleportation operation is admissible only if the following conditions are met:  
\begin{enumerate*}
    \item The two communication qubits must be \textit{free}, i.e., no logical qubits involved in the circuit should be mapped onto them.
    \item The quantum state to be teleported must reside in a physical qubit adjacent to the mediator communication qubit.
    \item The destination core must have sufficient available capacity.
\end{enumerate*}  

Thus, the extended energy calculation in \brand{} must account not only for the distance between logical qubits but also for the progress toward teleportation operations that can reduce this distance. Additional SWAP operations may be necessary to free the communication qubits required for teleportation. Moreover, an inter-core routing mechanism is required to establish a viable path between the cores of the two interacting qubits. This routing process must consider the current capacity of each core, ensuring that an inter-core path does not pass through a full core. A core is considered full if it has only one free physical qubit remaining.  

To prevent deadlock situations, layouts in which a core has no free physical qubits are disallowed, as such a configuration would prevent freeing the communication qubit via SWAP operations, thereby making qubit transfer out of the core impossible. If multiple communication qubits per core are available, the inter-core routing strategy should also account for their relative distances from the logical qubits involved in the gate operation. Additionally, certain communication qubits, despite being physically closer, may still be occupied by a logical qubit, making entanglement infeasible until further SWAP operations are performed to free them.  

For \brand{} to be effective, the routing process must incorporate all these factors, and the energy function should be designed to reflect these considerations. 
We achieve this by performing routing on a contracted graph, where edge weights are adjusted to account for the swap (hop) distance (precomputed in $\mathbf{D}$) between physical qubits, the proximity of the nearest free qubit to the communication qubits, and penalties associated with cores that have depleted capacity. The process of constructing the contracted graph and performing routing is illustrated in Fig.~\ref{fig:contracted}.  The procedure begins by identifying the physical qubits corresponding to the two disjoint logical qubits involved in the same gate operation, nearest free phyisical qubits to each communication qubit and their distance as highlighted by red dashed arrows in Fig.~\ref{fig:contracted}a. Next, the contracted graph is constructed by retaining only the communication qubits and the two qubits involved in front gate under consideration, with edge weights set to the shortest path length between them (Fig.~\ref{fig:contracted}b). For edges connecting the phyisical qubits containing the interacting logical qubits to the communications one is subtracted to the edge weight the absolute value is taken to make into account that the weight is zero when the interaacting qubits are adjacent to the communication qubits and not overlapping. To account for the accessibility of communication qubits, half of the hop distance to the nearest free qubit is added to the weights of the two edges incident to each communication qubit (Fig.~\ref{fig:contracted}c). This adjustment reflects the number of SWAP operations required to free a communication qubit if it is part of the shortest path.  Furthermore, if a core has exhausted its capacity (i.e., fewer than two free qubits remain), an additional penalty is incorporated into the edge weights of communication qubits associated with that core.

Finally, as shown in Fig.~\ref{fig:contracted}d, Dijkstra's algorithm \cite{dijkstra2022note} is applied to determine the shortest path length (\ie, sum of the weights of the shortest path edges) $\text{routing}_\lambda(g)$ between the two disjoint gate qubits $(q_{g,1},q_{g,2})$, which represents the minimum number of operations required to execute the gate. When multiple gates require inter-core routing in the same \brand{} iteration, traffic on inter-core links connecting two communication qubits on different core, can be considered by adding a traffic coefficient to the corresponding contracted graph edges before performing the routing for the next front gate.
The example shown in Fig.~\ref{fig:contracted} shows a trivial situation in which finding the shortest path is naive as one of the two intermediate cores is full and each core has only two communication qubits, hence no intra-core routing between communication qubits is needed.

Hence, the \brand{} total state energy can be still defined as in Eq.~\ref{eq:sabre}, but the definition of the gate energy is updated as follows:
\begin{equation*}
\text{energy}^G_{\lambda}(g) =
\begin{cases}
    \dist_\lambda(g), & \text{if } \text{core}_\lambda(q_{g,1}) = \text{core}_\lambda(q_{g,2}) \\
    \text{routing}_{\lambda}(g), & \text{otherwise}
\end{cases}
\end{equation*}

\subsection{Candidate Operations Selection}
\label{sec:candidates}
One of the key strengths of the SABRE algorithm is its ability to avoid evaluating all possible SWAP operations that could be applied to the layout. Instead, it restricts the set of candidate SWAPs at each iteration to those involving physical qubits where the logical qubits of gates in the current front are currently allocated.  

We adopt a similar approach but introduce modifications to accommodate logical qubit teleportation and gate teleportation. Specifically, the set of candidate SWAPs must also include those involving the nearest free qubits to communication qubits that participate in the routing plans determined by Dijkstra's algorithm for gates in the front layer where the two logical qubits reside in different cores. In addition to SWAP operations, the set of candidate operations must also incorporate feasible teleportation and telegate operations based on the current layout.  

\subsection{Initial Assignment}
\label{sec:initial}
The initial layout is generated by prioritizing the placement of interacting qubits from the first frontier of the circuit's DAG within the same cores, while intra-core positioning is randomly assigned. This initial configuration can serve as a starting point for further initial layout optimization, following an approach analogous to the original SABRE methodology. To obtain an optimized initial layout, we execute the layout synthesis process in three phases: first, a forward pass is performed, after which the resulting final layout is utilized for a second backward pass on the reversed quantum circuit. The final layout obtained from this backward pass then serves as the initial layout for the definitive forward pass, completing the bidirectional optimization procedure.

\subsection{Algorithmic Complexity}
\label{sec:complexity}

The original SABRE algorithm achieves a significant reduction in computational complexity by restricting the search space for SWAP operations, reducing it from $O(\exp(N))$ to $O(N)$. In the worst case, where all qubits are involved in the front layer, this approach remains scalable. Although additional SWAP operations may be required to bring qubits together, the maximum number of SWAPs needed for a two-qubit gate is bounded by the diameter of the chip coupling graph, which is $O(\sqrt{N})$ for a two-dimensional layout. Furthermore, evaluating the energy function is $O(N)$\footnote{This can be reduced to $O(1)$ with an optimized implementation \cite{zou2024lightsabre}}. As a result, the overall complexity for each two-qubit gate is reduced to at most $O(N^{2.5})$ \cite{sabre}. 

In \brand{}, when two qubits involved in a gate reside in different cores, routing decisions require executing Dijkstra's algorithm on a contracted graph that consists of communication qubits. The number of communication qubits is proportional to the number of cores, $C$. Since Dijkstra's algorithm runs in $O(V \log V + E)$ time for a graph with $V$ nodes and $E$ edges (using a priority queue implementation), and the number of communication qubits is $O(C)$, the additional computational overhead for inter-core routing is $O(C \log C)$. In the worst-case scenario, where every two-qubit gate requires inter-core routing, the total complexity per two-qubit gate increases to $O(N^{2.5} \cdot C \log C)$. This result indicates that while our approach introduces additional computational overhead due to inter-core routing, it remains scalable when $C$ is moderate compared to $N$. 
In this analysis, we assumed that finding the nearest free qubits to the communication qubits is $O(1)$ because for each communication qubit a sparse bucket queue can be used to keep track of the nearest free qubits as swaps and teleport are applied on them. Optimized implementations considering the calculation of the energy difference to greatly reduce complexity could be investigated in future research \cite{zou2024lightsabre}.

\section{Experiments}
\label{sec:experiments}

In this section we evaluate the proposed methodology, comparing it against state-of-the-art methods and near-optimal methods and evaluating the impact of optimizing the initial layout.

\subsection{Experimental Setup}

\begin{figure}
    \centering
    \includegraphics[width=\linewidth]{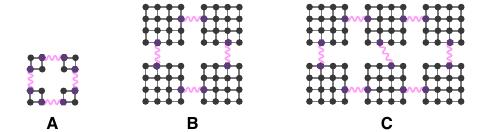}
    \caption{Architectures considered in experiments}
    \label{fig:archs}
\end{figure}


We implemented the proposed methodology using Python. To evaluate the performance of \brand{}, we considered the multi-core architectures illustrated in Fig.~\ref{fig:archs}, with the number of physical qubits ranging from 8 to 96. Our evaluation utilized a diverse set of benchmark circuits from the Qiskit framework~\cite{qiskit2024} and MQTBench~\cite{quetschlich2023mqtbench}, including: Random circuit, Greenberger–Horne–Zeilinger state preparation (GHZ), Graph State preparation, Portfolio Optimization with Quantum Approximate Optimization Algorithm (Portfolio QAOA), Amplitude Estimation (AE), Quantum Fourier Transform (QFT), Quantum Neural Network (QNN), Draper Adder, and Cuccaro Adder. All circuits were optimized, transpiled, and decomposed using Qiskit to target a native gate set consisting of rotation around z-axis (RZ), square root of NOT (SX), bit-flip (X), and controlled bit-flip (CX) operations. The experiments were conducted with varying numbers of logical qubits (8, 26, or 64). Number of introduced teleport, telegate and swap operations and compiled circuit depth were considered as evaluation metric.

\subsection{Comparison against state-of-the-art}

\begin{figure}
    \centering
    \includegraphics[width=\linewidth]{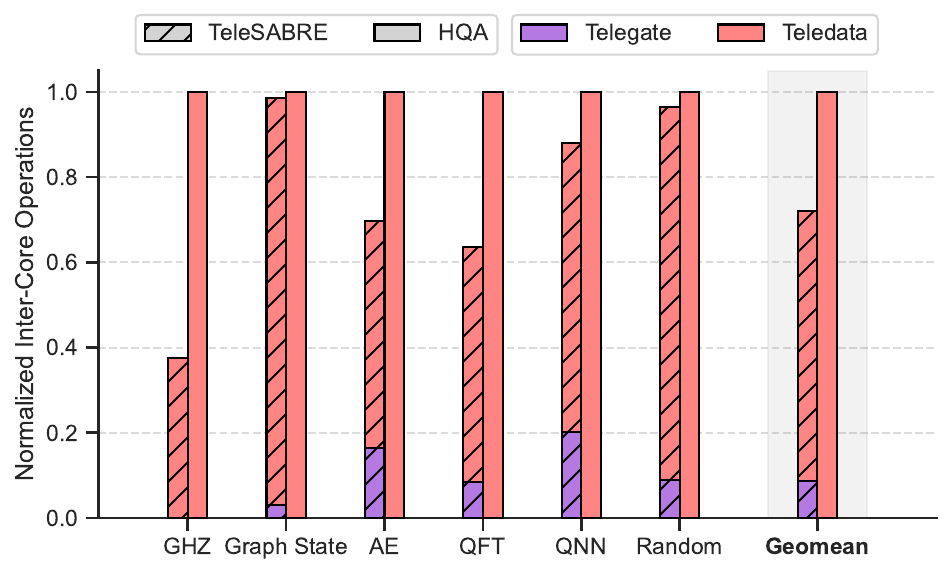}
    \caption{Inter-core operation count comparison against Hungarian Qubit Assignement for benchmark circuits with 64 qubits on architecture C.}
    \label{fig:hungarian}
\end{figure}

In this section we compare against a state-of-the-art method for inter-core state transfer minimization in multi-core quantum architectures, namely Hungarian Qubit Assignment (HQA)~\cite{escofet2023hungarian}. We consider the same initial qubit assignment in cores for both the methods and compare the amount of introduced inter-core operations, either state transfer (teledata) or remote gates (telegate). We consider architecture C with 96 physical qubits (Fig.~\ref{fig:archs}c) and benchmark circuits with 64 qubits. The results are shown in Fig.~\ref{fig:hungarian}. Except for the Graph State preparation circuit and the Random circuit, \brand{} achieves reductions in terms of inter-core operations ranging from $11.8\%$ to $62.5\%$. The geometric mean of reduction across all the benchmarks circuit is $28\%$.

\subsection{Comparison against near-optimal}

\begin{figure}
    \centering
    \includegraphics[width=\linewidth]{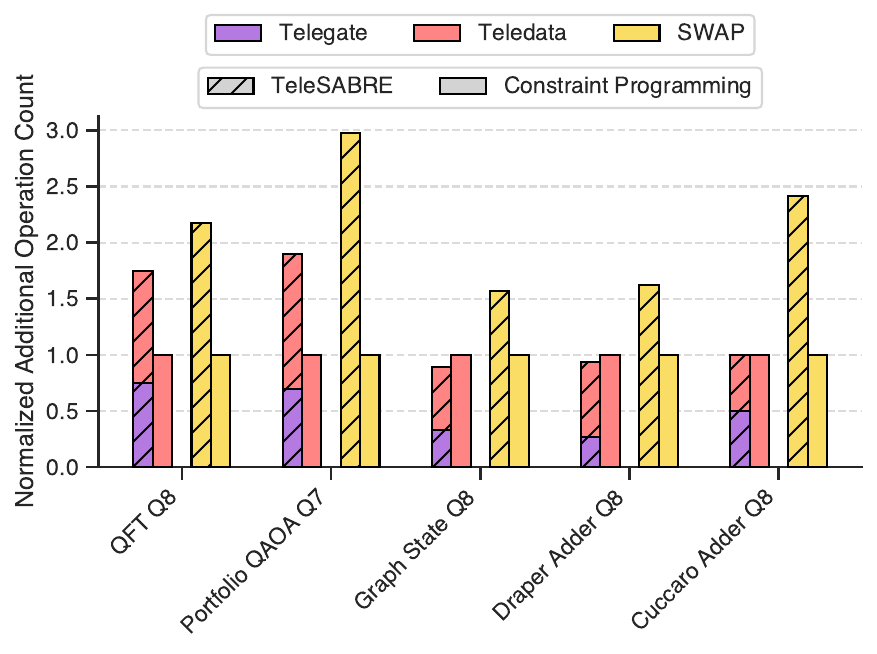}
    \caption{Operations introduced by layout synthesis compared to near-optimal constraint programming solutions for circuits up to 8 qubits on architecture A.}
    \label{fig:vs_cp}
\end{figure}

In this section, we evaluate the optimality of solutions generated by \brand{}. We employ a block-based constraint programming formulation for the layout synthesis problem in multi-core quantum systems to find near-optimal solutions. Our approach extends \mbox{TB-OLSQ2}~\cite{lin2023scalable} by accounting for intra-core swaps, inter-core teleports, and communication qubit logistics. We implement our model using OR-Tools~\cite{ortools}. The optimization follows a hierarchical objective: first minimizing transitions~\cite{lin2023scalable}, then teleports, and finally swaps. As solving the layout synthesis problem to optimality is NP-complete~\cite{botea2018complexity}, we are able to find near-optimal solutions only for small circuit instances and a small architecture, namely architecture A in Fig.~\ref{fig:archs}. Due to the stochasticity of the proposed algorithm, we run \brand{} 3 times and consider the best solution in terms of the count of intercore operations. Fig.~\ref{fig:vs_cp} shows the amount of inter-core and intra-core operations introduced by the two compilation methods. \brand{} finds solution with up to three times the near-optimal amount of swaps and two times the near-optimal amount of teleports. On the smaller A architecture we also notice an increase in the amount of telegates with respect to teledata operations. In fact, in bigger architectures more hops (teledata) are necessary on average to obtain two interating qubits on adjacent cores.

\subsection{Initial Layout Optimization}

\begin{figure}
    \centering
    \includegraphics[width=\linewidth]{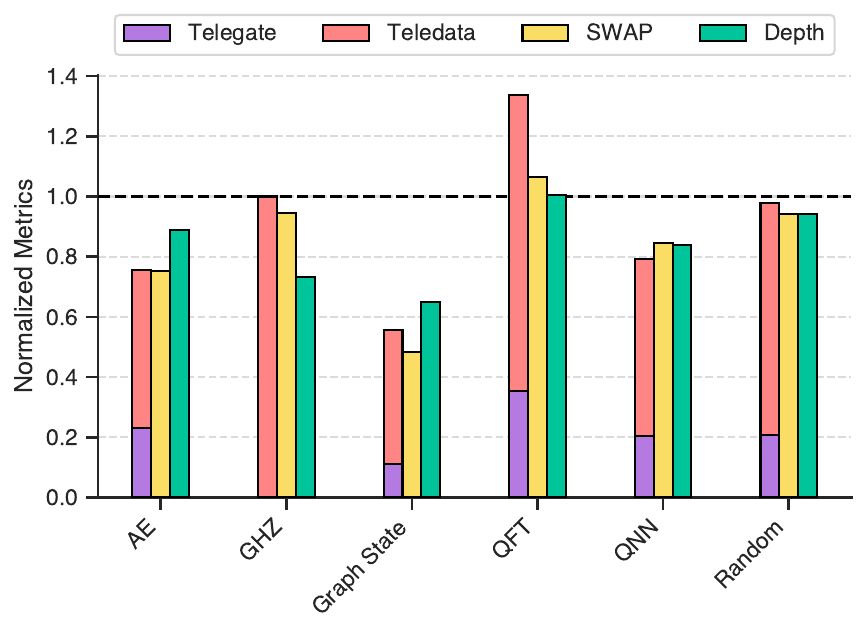}
    \caption{Normalized metrics comparison with respect to non-optimized initial layout.}
    \label{fig:initial}
\end{figure}

As described in Section~\ref{sec:initial}, it is also possible to further optimize the initial layout similarly to the original SABRE implementation. In this section, we evaluate the improvements in terms of different metrics with respect to the selection of the initial layout. We consideredhitecture B (Fig.~\ref{fig:archs}b) and 25 logical qubit circuits. Except for the QFT benchmark in which running the two extra forward-backward passes result in a worse initial layout, for all the other benchmarks the improvements are made up to $44\%$ in terms of inter-core operations, up to $51\%$ in terms of additional intra-core swap operations and up to $35\%$ in terms of circuit depth.

\section{Limitations and future work}
\label{sec:limitations}

\begin{figure}
    \centering
    \includegraphics{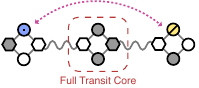}
    \caption{Example deadlock scenario.}
    \label{fig:deadlock}
\end{figure}

The proposed approach provides a heuristic baseline for holistic quantum compilation in multi-core architectures with teleport interconnects. In this section, we describe limitations that could be addressed in future work.

A first limitation concerns the possibility of encountering deadlocks during the layout synthesis process. For instance, consider the scenario in Fig.~\ref{fig:deadlock}, where the Dijkstra routing phase fails to find a feasible teleport path between two cores because the only possible path contains a core with depleted capacity. In this case, the only viable operation is to perform a teleport to free a spot in the intermediate core. Deadlock situations are also present in the original SABRE algorithm and have been solved through a safety valve mechanism~\cite{zou2024lightsabre}; a similar approach could be explored in future research for \brand{}.

Another limitation is the lack of exploitation of interesting mechanisms enabled by entanglement properties. In \brand{}, there is no tracking of entangled pairs, as it only keeps track of the possibility to create them when needed for teleport operations between directly connected cores. However, Bell state qubits can be moved after their creation from the physical communication qubits where they are initially allocated. Furthermore, entanglement swapping can be used to perform teleportation between cores that are not directly connected via an inter-core coupling but have a path between them \cite{caleffi2024distributed, ferrari2021compiler}. 

Finally, similar to the first version of SABRE~\cite{sabre}, \brand{} was implemented in Python. Further effort should be devoted to developing an efficient implementation, as was done for the original SABRE algorithm in~\cite{zou2024lightsabre}. Furthermore, the Dijkstra algorithm could be replaced with the A* algorithm, using the number of cores with depleted capacity along a path as a heuristic to quickly prioritize paths with available resources.

\section{Related Works}
\label{sec:related_works}

Several works have addressed the quantum layout synthesis problem for single-core architectures with optimal and heuristic approaches  \cite{botea2018complexity, siraichi2018qubit, lin2023scalable, li2019tackling, nannicini2022optimal, kremer2024practical, huang2022reinforcement} optimizing the number of intra-core swap operations and compiled circuit depth. Recently, many works focused on multi-core, modular and distributed architectures. Many of these methods focused on minimizing inter-core state transfers without considering the challenges of the teleportation protocol and the intra-core topologies \cite{andres2019automated, daei2020optimized, baker2020time, g2021efficient, escofet2023hungarian, bandic2023mapping, cuomo2023optimized, escofet2024revisiting, russo2024attention}. Others considered inter-core communication primitives different from teleport protocols \cite{zhang2023compilation,escofet2024route}. Leveraging teleport operations has also been proposed as a way to reduce swap overheads in single-core architectures \cite{hillmich2021exploiting,padda2024improving}. Due to combinatorial nature of the problem, reinforcement learning is also being investigated as a potential way to derive heuristics \cite{promponas2024compiler,huang2022reinforcement,russo2024attention}. In \cite{ferrari2021compiler} the authors propose a compiler for nearest neighbour distributed topologies taking into account entanglement swapping. In \cite{ferrari2023modular} the authors adopt a two-level optimization approach by first performing qubit assignement based on k-cut partitioning.

\section{Conclusion}
\label{sec:conclusions}

We have presented an extension of the SABRE algorithm that offers a holistic approach to quantum layout synthesis in multi-core architectures with teleport-based interconnects. Future research should focus on enhancing algorithm efficiency, examining how algorithm parameters influence the four critical output metrics, \ie amount of teledata, telegate, swap operations, and circuit depth, and assessing their impact on circuit fidelity~\cite{hopf2025improving}. Additional important directions include implementing deadlock prevention mechanisms and exploring alternative energy formulations. We believe that our publicly available implementation of \brand{} could serve as a valuable baseline for future heuristic algorithms in this domain.

\section*{Acknowledgement}
Authors gratefully acknowledge funding from the European Commission through HORIZON-EIC-2022-PATHFINDEROPEN01-101099697 (QUADRATURE)

\bibliographystyle{bib/IEEETran}
\bibliography{bib/IEEEabrv,references}

\begin{thebibliography}{10}
\providecommand{\url}[1]{#1}
\csname url@samestyle\endcsname
\providecommand{\newblock}{\relax}
\providecommand{\bibinfo}[2]{#2}
\providecommand{\BIBentrySTDinterwordspacing}{\spaceskip=0pt\relax}
\providecommand{\BIBentryALTinterwordstretchfactor}{4}
\providecommand{\BIBentryALTinterwordspacing}{\spaceskip=\fontdimen2\font plus
\BIBentryALTinterwordstretchfactor\fontdimen3\font minus \fontdimen4\font\relax}
\providecommand{\BIBforeignlanguage}[2]{{%
\expandafter\ifx\csname l@#1\endcsname\relax
\typeout{** WARNING: IEEEtran.bst: No hyphenation pattern has been}%
\typeout{** loaded for the language `#1'. Using the pattern for}%
\typeout{** the default language instead.}%
\else
\language=\csname l@#1\endcsname
\fi
#2}}
\providecommand{\BIBdecl}{\relax}
\BIBdecl

\bibitem{montanaro2016quantum}
A.~Montanaro, ``Quantum algorithms: an overview,'' \emph{npj Quantum Information}, vol.~2, no.~1, pp. 1--8, 2016.

\bibitem{shor1999polynomial}
P.~W. Shor, ``Polynomial-time algorithms for prime factorization and discrete logarithms on a quantum computer,'' \emph{SIAM review}, vol.~41, no.~2, pp. 303--332, 1999.

\bibitem{grover1996fast}
L.~K. Grover, ``A fast quantum mechanical algorithm for database search,'' in \emph{ACM symposium on Theory of computing}, 1996, pp. 212--219.

\bibitem{nielsen2010quantum}
M.~A. Nielsen and I.~L. Chuang, \emph{Quantum computation and quantum information}.\hskip 1em plus 0.5em minus 0.4em\relax Cambridge university press, 2010.

\bibitem{preskill2018quantum}
J.~Preskill, ``Quantum computing in the nisq era and beyond,'' \emph{Quantum}, vol.~2, p.~79, 2018.

\bibitem{aaronson2013quantum}
S.~Aaronson, \emph{Quantum computing since Democritus}.\hskip 1em plus 0.5em minus 0.4em\relax Cambridge University Press, 2013.

\bibitem{jnane2022multicore}
H.~Jnane, B.~Undseth, Z.~Cai, S.~C. Benjamin, and B.~Koczor, ``Multicore quantum computing,'' \emph{Physical Review Applied}, vol.~18, no.~4, 2022.

\bibitem{xie2021mitigating}
L.~Xie, J.~Zhai, and W.~Zheng, ``Mitigating crosstalk in quantum computers through commutativity-based instruction reordering,'' in \emph{2021 58th ACM/IEEE Design Automation Conference (DAC)}.\hskip 1em plus 0.5em minus 0.4em\relax IEEE, 2021, pp. 445--450.

\bibitem{sarovar2020detecting}
M.~Sarovar, T.~Proctor, K.~Rudinger, K.~Young, E.~Nielsen, and R.~Blume-Kohout, ``Detecting crosstalk errors in quantum information processors,'' \emph{Quantum}, vol.~4, p. 321, 2020.

\bibitem{arute2019quantum}
F.~Arute, K.~Arya, R.~Babbush, D.~Bacon, J.~C. Bardin, R.~Barends, R.~Biswas, S.~Boixo, F.~G. Brandao, D.~A. Buell \emph{et~al.}, ``Quantum supremacy using a programmable superconducting processor,'' \emph{Nature}, vol. 574, no. 7779, pp. 505--510, 2019.

\bibitem{smith2022scaling}
K.~N. Smith, G.~S. Ravi, J.~M. Baker, and F.~T. Chong, ``Scaling superconducting quantum computers with chiplet architectures,'' in \emph{IEEE/ACM International Symposium on Microarchitecture (MICRO)}, 2022, pp. 1092--1109.

\bibitem{alarcorn_iscas23}
E.~Alarcón, S.~Abadal, F.~Sebastiano, M.~Babaie, E.~Charbon, P.~H. Bolívar, M.~Palesi, E.~Blokhina, D.~Leipold, B.~Staszewski, A.~Garcia-Sáez, and C.~G. Almudever, ``Scalable multi-chip quantum architectures enabled by cryogenic hybrid wireless/quantum-coherent network-in-package,'' in \emph{2023 IEEE International Symposium on Circuits and Systems (ISCAS)}, 2023, pp. 1--5.

\bibitem{botea2018complexity}
A.~Botea, A.~Kishimoto, and R.~Marinescu, ``On the complexity of quantum circuit compilation,'' in \emph{Proceedings of the International Symposium on Combinatorial Search}, vol.~9, no.~1, 2018, pp. 138--142.

\bibitem{siraichi2018qubit}
M.~Y. Siraichi, V.~F.~d. Santos, C.~Collange, and F.~M.~Q. Pereira, ``Qubit allocation,'' in \emph{International Symposium on Code Generation and Optimization}, 2018, pp. 113--125.

\bibitem{sabre}
G.~Li, Y.~Ding, and Y.~Xie, ``Tackling the qubit mapping problem for nisq-era quantum devices,'' in \emph{Proceedings of the twenty-fourth international conference on architectural support for programming languages and operating systems}, 2019, pp. 1001--1014.

\bibitem{caleffi2024distributed}
M.~Caleffi, M.~Amoretti, D.~Ferrari, J.~Illiano, A.~Manzalini, and A.~S. Cacciapuoti, ``Distributed quantum computing: a survey,'' \emph{Computer Networks}, vol. 254, p. 110672, 2024.

\bibitem{rodrigo2021modelling}
S.~Rodrigo, S.~Abadal, C.~G. Almud{\'e}ver, and E.~Alarc{\'o}n, ``Modelling short-range quantum teleportation for scalable multi-core quantum computing architectures,'' in \emph{ACM International Conference on Nanoscale Computing and Communication}, 2021, pp. 1--7.

\bibitem{zou2024lightsabre}
H.~Zou, M.~Treinish, K.~Hartman, A.~Ivrii, and J.~Lishman, ``Lightsabre: A lightweight and enhanced sabre algorithm,'' \emph{arXiv preprint arXiv:2409.08368}, 2024.

\bibitem{dijkstra2022note}
E.~W. Dijkstra, ``A note on two problems in connexion with graphs,'' in \emph{Edsger Wybe Dijkstra: his life, work, and legacy}, 2022, pp. 287--290.

\bibitem{qiskit2024}
A.~Javadi-Abhari, M.~Treinish, K.~Krsulich, C.~J. Wood, J.~Lishman, J.~Gacon, S.~Martiel, P.~D. Nation, L.~S. Bishop, A.~W. Cross, B.~R. Johnson, and J.~M. Gambetta, ``Quantum computing with {Q}iskit,'' 2024.

\bibitem{quetschlich2023mqtbench}
N.~Quetschlich, L.~Burgholzer, and R.~Wille, ``{{MQT Bench}}: Benchmarking software and design automation tools for quantum computing,'' \emph{{Quantum}}, 2023, {{MQT Bench}} is available at \url{https://www.cda.cit.tum.de/mqtbench/}.

\bibitem{escofet2023hungarian}
P.~Escofet, A.~Ovide, C.~G. Almudever, E.~Alarc{\'o}n, and S.~Abadal, ``Hungarian qubit assignment for optimized mapping of quantum circuits on multi-core architectures,'' \emph{IEEE Computer Architecture Letters}, vol.~22, no.~2, pp. 161--164, 2023.

\bibitem{lin2023scalable}
W.-H. Lin, J.~Kimko, B.~Tan, N.~Bj{\o}rner, and J.~Cong, ``Scalable optimal layout synthesis for nisq quantum processors,'' in \emph{2023 60th ACM/IEEE Design Automation Conference (DAC)}.\hskip 1em plus 0.5em minus 0.4em\relax IEEE, 2023, pp. 1--6.

\bibitem{ortools}
\BIBentryALTinterwordspacing
L.~Perron and V.~Furnon, ``Or-tools,'' Google. [Online]. Available: \url{https://developers.google.com/optimization/}
\BIBentrySTDinterwordspacing

\bibitem{ferrari2021compiler}
D.~Ferrari, A.~S. Cacciapuoti, M.~Amoretti, and M.~Caleffi, ``Compiler design for distributed quantum computing,'' \emph{IEEE Transactions on Quantum Engineering}, vol.~2, pp. 1--20, 2021.

\bibitem{li2019tackling}
G.~Li, Y.~Ding, and Y.~Xie, ``Tackling the qubit mapping problem for nisq-era quantum devices,'' in \emph{International Conference on Architectural Support for Programming Languages and Operating Systems}, 2019, pp. 1001--1014.

\bibitem{nannicini2022optimal}
G.~Nannicini, L.~S. Bishop, O.~G{\"u}nl{\"u}k, and P.~Jurcevic, ``Optimal qubit assignment and routing via integer programming,'' \emph{ACM Transactions on Quantum Computing}, vol.~4, no.~1, pp. 1--31, 2022.

\bibitem{kremer2024practical}
D.~Kremer, V.~Villar, H.~Paik, I.~Duran, I.~Faro, and J.~Cruz-Benito, ``Practical and efficient quantum circuit synthesis and transpiling with reinforcement learning,'' \emph{arXiv preprint arXiv:2405.13196}, 2024.

\bibitem{huang2022reinforcement}
C.-Y. Huang, C.-H. Lien, and W.-K. Mak, ``Reinforcement learning and dear framework for solving the qubit mapping problem,'' in \emph{41st IEEE/ACM international conference on computer-aided design}, 2022, pp. 1--9.

\bibitem{andres2019automated}
P.~Andres-Martinez and C.~Heunen, ``Automated distribution of quantum circuits via hypergraph partitioning,'' \emph{Physical Review A}, vol. 100, no.~3, p. 032308, 2019.

\bibitem{daei2020optimized}
O.~Daei, K.~Navi, and M.~Zomorodi-Moghadam, ``Optimized quantum circuit partitioning,'' \emph{International Journal of Theoretical Physics}, vol.~59, no.~12, pp. 3804--3820, 2020.

\bibitem{baker2020time}
J.~M. Baker, C.~Duckering, A.~Hoover, and F.~T. Chong, ``Time-sliced quantum circuit partitioning for modular architectures,'' in \emph{Proceedings of the 17th ACM International Conference on Computing Frontiers}, 2020, pp. 98--107.

\bibitem{g2021efficient}
R.~G~Sundaram, H.~Gupta, and C.~Ramakrishnan, ``Efficient distribution of quantum circuits,'' in \emph{35th International Symposium on Distributed Computing (DISC 2021)}.\hskip 1em plus 0.5em minus 0.4em\relax Schloss Dagstuhl--Leibniz-Zentrum f{\"u}r Informatik, 2021, pp. 41--1.

\bibitem{bandic2023mapping}
M.~Bandic, L.~Prielinger, J.~N{\"u}{\ss}lein, A.~Ovide, S.~Rodrigo, S.~Abadal, H.~Van~Someren, G.~Vardoyan, E.~Alarcon, C.~G. Almudever \emph{et~al.}, ``Mapping quantum circuits to modular architectures with qubo,'' in \emph{2023 IEEE International Conference on Quantum Computing and Engineering (QCE)}, vol.~1.\hskip 1em plus 0.5em minus 0.4em\relax IEEE, 2023, pp. 790--801.

\bibitem{cuomo2023optimized}
D.~Cuomo, M.~Caleffi, K.~Krsulich, F.~Tramonto, G.~Agliardi, E.~Prati, and A.~S. Cacciapuoti, ``Optimized compiler for distributed quantum computing,'' \emph{ACM Transactions on Quantum Computing}, vol.~4, no.~2, pp. 1--29, 2023.

\bibitem{escofet2024revisiting}
P.~Escofet, A.~Ovide, M.~Bandic, L.~Prielinger, H.~van Someren, S.~Feld, E.~Alarc{\'o}n, S.~Abadal, and C.~G. Almud{\'e}ver, ``Revisiting the mapping of quantum circuits: Entering the multi-core era,'' \emph{ACM Transactions on Quantum Computing}, 2024.

\bibitem{russo2024attention}
E.~Russo, M.~Palesi, D.~Patti, G.~Ascia, and V.~Catania, ``Attention-based deep reinforcement learning for qubit allocation in modular quantum architectures,'' \emph{arXiv preprint arXiv:2406.11452}, 2024.

\bibitem{zhang2023compilation}
H.~Zhang, K.~Yin, A.~Wu, H.~Shapourian, A.~Shabani, and Y.~Ding, ``Mech: Multi-entry communication highway for superconducting quantum chiplets,'' in \emph{29th ACM International Conference on Architectural Support for Programming Languages and Operating Systems}, 2024.

\bibitem{escofet2024route}
P.~Escofet, A.~Gonzalvo, E.~Alarc{\'o}n, C.~G. Almud{\'e}ver, and S.~Abadal, ``Route-forcing: Scalable quantum circuit mapping for scalable quantum computing architectures,'' in \emph{2024 IEEE International Conference on Quantum Computing and Engineering (QCE)}, vol.~1.\hskip 1em plus 0.5em minus 0.4em\relax IEEE, 2024, pp. 909--920.

\bibitem{hillmich2021exploiting}
S.~Hillmich, A.~Zulehner, and R.~Wille, ``Exploiting quantum teleportation in quantum circuit mapping,'' in \emph{Proceedings of the 26th Asia and South Pacific Design Automation Conference}, 2021, pp. 792--797.

\bibitem{padda2024improving}
G.~Padda, E.~Tham, A.~Brodutch, and D.~Touchette, ``Improving qubit routing by using entanglement mediated remote gates,'' in \emph{2024 IEEE International Conference on Quantum Computing and Engineering (QCE)}, vol.~1.\hskip 1em plus 0.5em minus 0.4em\relax IEEE, 2024, pp. 1770--1776.

\bibitem{promponas2024compiler}
P.~Promponas, A.~Mudvari, L.~Della~Chiesa, P.~Polakos, L.~Samuel, and L.~Tassiulas, ``Compiler for distributed quantum computing: a reinforcement learning approach,'' \emph{arXiv preprint arXiv:2404.17077}, 2024.

\bibitem{ferrari2023modular}
D.~Ferrari, S.~Carretta, and M.~Amoretti, ``A modular quantum compilation framework for distributed quantum computing,'' \emph{IEEE Transactions on Quantum Engineering}, vol.~4, pp. 1--13, 2023.

\bibitem{hopf2025improving}
P.~Hopf, N.~Quetschlich, L.~Schulz, and R.~Wille, ``Improving figures of merit for quantum circuit compilation,'' \emph{arXiv preprint arXiv:2501.13155}, 2025.

\end{thebibliography}

\end{document}